\documentclass[12pt,a4paper]{article}
\usepackage[utf8]{inputenc}
\usepackage{amsmath}
\usepackage{amsfonts}
\usepackage{amssymb}
\usepackage{hyperref}
\usepackage{graphicx}
\usepackage{lmodern}
\usepackage[left=2cm,right=2cm,top=2cm,bottom=2cm]{geometry}
\title{Comparative Study of $\alpha$ - $\alpha$ interaction potentials constructed using various phenomenological models }
\author{Ayushi Awasthi  and O.S.K.S Sastri}
\begin{document}
\maketitle
\begin{abstract}
In this paper, we have made a comparative study of $\alpha-\alpha$ scattering using different phenomenological models like Morse, double gaussian, double Hulthén, Malfliet-Tjon and double exponential for the nuclear interaction and atomic Hulthén as screened coulomb potential. The phase equations for S, D and G channels have been numerically solved using $5^{th }$ order Runge-Kutta Method to compute scattering phase shifts (SPS) for elastic scattering region consisting of eneries up to 25.5 MeV. The model parameters in each of the chosen potentials were varied in an iterative fashion to minimize the mean absolute percentage error (MAPE) between simulated and expected SPS. A comparative analysis revealed that, all the phenomenological models result in exactly similar inverse potentials with closely matching MAPE values for S , D and G state. One can conclude that any mathematical function that can capture the basic features of two body interaction would always guide correctly in construction of inverse potentials.
 
\vspace{0.5cm}


\end{abstract} 

\section{Introduction}
Scattering studies of $\alpha$ particles with $^{4}_{2}He$ nuclei is of importance in understanding nature of nuclear force and also for gaining insights into few body \cite{1} and cluster models \cite{2,3}. Rutherford $\&$ Chadwick were the first to study $\alpha$- $\alpha$ scattering in 1927 \cite{4} and since then, numerous experiments have been performed at various energy levels to deepen our understanding. In 1956, Heydenburg and Temmer presented experimental scattering phase shifts (SPS) for the low-energy range of 0.6 MeV to 3 MeV \cite{5}. Tombrello and Senhouse, in 1963, provided experimental SPS covering the energy range of 3.84 MeV to 11.88 MeV \cite{6}. Then, SPS for energies between 12.3 MeV and 22.9 MeV were given by Nilson et al. in 1958 \cite{7}. Subsequently, Chien and Brown, in 1974, contributed experimental SPS for the energy range of 18 MeV to 29.50 MeV \cite{8}.

The SPS data obtained from these experiments were compiled by Afzal et al.\cite{9} which is generally considered by theoretical physicists for studying $\alpha - \alpha$ scattering. However, it is worth noting that their compilation included data only up until 1969. Recognizing the significance of incorporating Chien and Brown data from 1974, Anil et. al. took the initiative to update the database for $\alpha$-$\alpha$ scattering in 2022 \cite{10}.
 
In the realm of theoretical physics, numerous phenomenological models have emerged and evolved over the past six decades. Notably, in 1964, Darriulat et al. \cite{11} embarked upon a significant endeavor by employing the Woods-Saxon potential within an optical model. Their objective was to extract SPS for various angular momentum states, specifically $\ell$ = 0, 2, 4, 6 and 8, spanning an energy range between 53 MeV and 120 MeV. %, discussing inelastic scattering.

Almost at the same time, Ali and Bodmer ventured into the study of $\alpha$-$\alpha$ scattering \cite{12}. In their investigation, they employed a Double Gaussian potential with four adjustable parameters. Their approach involved an initial determination of the attractive component of the nuclear force by fitting the available scattering data in the $\ell$ = 4 channel. Then constraining the shape of potential for large distances, they obtained the repulsive nature exhibited in the $\ell$ = 0 and $\ell$ = 2 channels, at short distances. 
 
In 1977, Buck et al. \cite{13} put forth a compelling argument, emphasizing that local potential is sufficient to model the interaction between $\alpha$ particles, by meticulously examining of two notable models: the Resonating Group Method (RGM)\cite{14} and the Orthogonality Condition Model (OCM)\cite{15}. They employed a single Gaussian function characterized by two parameters. These are obtained by selecting experimental energy value of pseudo-bound scattering state E= 0.0198 MeV and phase shift for $\ell$ = 2, at 3 MeV. They  were able to provide a reasonable explanation of the observed SPS for $\ell$ = $0$, $2$, $4$ and $6$ for energy values upto $E_{\ell ab}$ = 80 MeV.

 In 2003, M. Odsuren et.al combined two approaches, Complex Scaling Method(CSM) and Orthogonality Condition Method(OCM) called CSOCM \cite{16} to compute resonance states in two-body systems, by including the influence of Pauli exclusion principle between clusters. They have applied two different potentials, Gaussian and harmonic oscillator, and obtained wave functions for $\alpha$ - $\alpha$
system to calculate resonance energies with their decay
widths. During their calculations, they considered SPS for partial waves $\ell$ = 0, 2, 4, 6 and 8, with energies upto 50 MeV.

Recently, Anil et.al \cite{17} have revisited the local gaussian potential with an innovative algorithm that is a combination of the Matrix Method \cite{18} and Variational Monte Carlo (VMC)\cite{19} technique. In this approach, they have considered the bound state energies as given in Buck et.al \cite{13} to optimize the model parameters and then utilising the determined interaction potential in phase function method (PFM), they have obtained SPS for $\ell$ = 0, 2 $\&$ 4 channels for energies upto 25.5 MeV. Then, they proposed Morse potential as nuclear interaction and directly utilized all available experimental SPS for optimising the model parameters. This is akin to constructing the model from the data, as in machine learning paradigm, which is fundamentally the approach of inverse scattering theory \cite{20}. All these above procedures \cite{10,12,13,16,17} utilised \textit{erf()} function based coulomb interaction.
 
Alternatively, Laha et al.\cite{21,22,23} have utilized PFM to calculate SPS and obtain interaction potentials. Thay employed the double Hulthén potential to describe the nuclear interaction, while adopting the atomic Hulthén ansatz to account for the screened Coulomb interaction \cite{24}. Their noteworthy study focused on investigating $\alpha$-$\alpha$ scattering up to an  energy range of $E_{\ell ab} = 100$ MeV. 
%to obtain better convergence with expected data and compiled an updated database for $\alpha$ - $\alpha$ scattering \cite{10}. 
%  
The motivation behind this study was based on the following observations:
 
Firstly, we have observed that for Morse + \textit{erf()} ansatz of Anil et.al.\cite{10}, the depth of the potential for $\ell$ = 2 is not shallower than that of $\ell$ = 0. Therefore, we became intrigued to consider the performance of the atomic Hulthén screening potential as a replacement for \textit{erf()}. This, in turn, led us to include a similar study for the double Gaussian potential \cite{12}.

Secondly, we observed that there were three studies\cite{20,21,22} on $\alpha$-$\alpha$ scattering using the Double Hulthén potential as the nuclear interaction, with different screening radii. However, with those potential parameters, the height of the Coulomb barrier for $\ell$ = 2 and 4 was not observed to be near their corresponding resonance energies \cite{25}. Therefore, we have opted to re-optimize the model parameters using our innovative algorithm within the elastic region, specifically up to 25.5 MeV. 

Thirdly, the Malfliet-Tjon (MT) potential \cite{26}, which is a combination of attractive and repulsive forms of the Yukawa potential \cite{27}, has been able to reasonably explain the SPS for n-p, n-d, and p-d systems \cite{28,29}. Therefore, we have incorporated this interaction potential for the first time in order to study $\alpha$-$\alpha$ scattering.

Finally, an observation that Morse potential is a composite of exponential functions has led us to incorporate the double exponential function into our analysis for the purpose of comparison. 

So, in this paper, our aim is to perform a comprehensive comparative analysis of various phenomenological potential models as local potentials to model the nuclear interaction between two alpha particles, namely Morse, Double Gaussian, Double Hulthén, Malfliet - Tjon (MT), and Double Exponential. Our study focuses on investigating the elastic scattering of alpha particles ( $\alpha$-$\alpha$) in the S, D and G channels, utilizing the atomic Hulthén potential as screened coulomb potential for energies ranging upto 25.5 MeV. 

\section{Methodology}
The interaction between two alpha particles is written as a combination of nuclear and Coulomb parts as
\begin{equation}
    V(r) = V_{N}(r) + V_{C}
\end{equation}
The nuclear part is modeled by various phenomenological potentials as follows:
\begin{itemize}
\item Morse Potential \cite{30}
\begin{equation}
V_{N}(r) = D_0\left(e^{-2(r-r_m)/a_m} - 2e^{-(r-r_m)/a_m}\right)
\label{Morse}
\end{equation}
where $D_{0}$ , $r_m$ and $a_m$ represent the depth of potential (in $fm^{-2}$), equilibrium distance (in $fm$) and shape of potential (in $fm$) respectively. It is a three parameter potential.
\item Double Gaussian Potential \cite{12}
\begin{equation}
V_{N}(r) = V_r e^{-\mu_r^2 r^2} -  V_a e^{-\mu_a^2 r^2}
\label{DG}
\end{equation}
where $V_r$ and $V_a$ represents the strength of repulsive and attractive parts in $fm^{-2}$, respectively, $\mu_r$ and $\mu_a$ are their corresponding inverse ranges in $fm^{-1}$. It is a four parameter potential.
 \item Double Hulthén Potential \cite{21}
\begin{equation}
V_{N}(r) = -S_{\ell_1}\frac{e^{-\beta r}}{(e^{-\alpha r} - e^{-\beta r})} + S_{\ell_2}\frac{e^{-(\beta+\alpha) r}}{(e^{-\alpha r} - e^{-\beta r})^2}
\label{DH}
\end{equation}
where $S_{\ell_1}$, $S_{\ell _2}$, $\alpha$ and $\beta$ are four parameters. The first two represent depth of potential (in $fm^{-2}$) and the rest two its range (in $fm^{-1}$) of potential.  
\item Malfliet-Tjon (MT) Potential \cite{26}
%\begin{equation}
%V_N(r) = \frac{V_r e^{-2 \alpha r} - V_a e^{-\alpha r}}{r}
%\label{MT}
%\end{equation}
%\begin{equation}
%V_N(r) = \frac{V_R e^{- \mu_{r} r} - V_A e^{-\mu r}}{r}
%\label{MT}
%\end{equation}
%where $\mu_{r}$ = 2$\mu$. 
%So, therefore eq.\ref{MT} becomes
\begin{equation}
   V_N(r) = \frac{V_R e^{- 2\mu r} - V_A e^{-\mu r}}{r} 
\end{equation}
where $V_R$ and $V_A$ represent depths of repulsive and attractive part of the potential in $fm^{-2}$  and $\mu $ is inverse range parameter in $fm^{-1}$.
\item Double Exponential
\begin{equation}
V_N(r) =  A e^{-\alpha_{1}r}- B e^{-\alpha_{2}r}
\label{DE}
\end{equation}
where $A$ and $B$ represent depths of repulsive and attractive part of the potential in $fm^{-2}$ and $\alpha_1 $ and $\alpha_2$ are inverse range parameters in $fm^{-1}$.
\end{itemize}
To account for Coulomb interaction, we consider the atomic Hulthén potential \cite{24} which is given as
\begin{equation}
V_{AH}(r)= V_o \frac{e^{-r/a}}{(1-e^{-r/a})}
\end{equation}
where $V_{o}$ is strength of the potential and a is the  screening radius. The two parameters $V_o$ and a are related by \cite{31}
\begin{equation*}
 V_o a = 2K\eta
\end{equation*}
where K is momentum energy in lab frame and $\eta$ is Sommerfeld parameter defined as
\begin{equation*}
\eta = \frac{\alpha}{\hbar v}
\end{equation*}
Here, $v$ is relative velocity of the reactants at large separation and $\alpha = Z_1Z_2e^2$.
So, 
\begin{equation*}
V_o a = \frac{Z_1 Z_2 e^2 \mu}{\hbar^2}
\end{equation*}
For $\alpha - \alpha$, $Z_1 = Z_2 = 2$ , $\mu = \frac{m_{\alpha}}{2} = 1864.38525$ $\frac{MeV}{c^2}$ , 
$e^2 = 1.44 MeV fm$ and therefore $V_o a = 0.2758 fm^{-1}$.

\subsection{Phase Function Method}
The time independent Schr$\ddot{o}$dinger equation (TISE) can be written as
\begin{equation}
\frac{d^2{u_{\ell}(r)}}{dr^2}+\bigg[k^2-\frac{\ell(\ell+1)}{r^2}-U(r)\bigg]u_{\ell}(r) = 0
\label{Scheq}
\end{equation}
where $U(r) = V(r)/(\hbar^2/2\mu)$ \& $k_{c.m} = \sqrt{E_{c.m}/(\hbar^2/2\mu)}$ and $E_{c.m} = 0.5 E_{\ell ab}$. 

For $\alpha-\alpha $ system, the value of $\hbar^2/2\mu$ = 10.44217 MeVfm$^{2}$.

Phase Function Method is one of the important tools in scattering studies for both local \cite{32} and non-local interactions \cite{33,34} .
The TISE in Eq.\ref{Scheq} can been transformed to a non-linear Riccati equation of first order\cite{32,36,36}, which directly deals with SPS information, given by: 
\begin{equation}
\delta_{\ell}'(k,r)=-\frac{U(r)}{k}\bigg[\cos(\delta_\ell(k,r))\hat{j}_{\ell}(kr)-\sin(\delta_\ell(k,r))\hat{\eta}_{\ell}(kr)\bigg]^2
\label{PFMeqn}
\end{equation}
The Riccati Hankel function of first kind is given by $\hat{h}_{\ell}(r)=-\hat{\eta}_{\ell}(r)+\textit{i}~ \hat{j}_{\ell}(r)$, where $\hat{j_{\ell}}(kr)$ is Ricatti-Bessel and $\hat{\eta_{\ell}}(kr)$ Riccati-Neumann function. By substituting the expressions for different $\ell$-values of these two later functions, we obtain the respective phase equations as: 
\begin{enumerate}
\item $\ell = 0$:
\begin{equation}
    \delta_0'(k,r)=-\frac{U(r)}{k}\sin^2[\delta_0 + \kappa]
\end{equation}
\item $\ell$ = 2:
\begin{equation}
\delta_2'(k,r) = -\frac{U(r)}{k}\bigg[-\sin{\left(\delta_2+ \kappa \right)}-\frac{3 \cos{\left(\delta_2 + \kappa \right)}}{\kappa} + \frac{3 \sin{\left(\delta_2 + \kappa \right)}}{\kappa^2}\bigg]^2 
\end{equation}
\item $\ell$ = 4
\begin{eqnarray}\nonumber
\delta_4'(k,r)=-\frac{U(r)}{k}\bigg[\sin{\left(\delta_4 + \kappa \right)} + \frac{10 \cos{\left(\delta_4 + \kappa \right)}}{\kappa}-\frac{45 \sin{\left(\delta_4 + \kappa \right)}}{ \kappa^2}\\~~~~~~~~~~~~~- \frac{105 \cos{\left(\delta_4 + \kappa \right)}}{ \kappa^3}+ \frac{105 \sin{\left(\delta_4 + \kappa \right)}}{ \kappa^4}\bigg]^2
\end{eqnarray}
\end{enumerate}
These equations are solved using 5th order Runge-Kutta methods by choosing the initial condition as $\delta_{\ell}(0,k) = 0$ and integrating to a large distance.

\section{Results and Discussion:}
The observed resonances in $\alpha-\alpha$ scattering experiments occurring at 0.09184 MeV, 3.03 MeV and 11.35 MeV \cite{37} corresponding to $\ell= 0$, $2$ and $4$ channels respectively provide an understanding of $^8Be$ nuclear structure. These are named as S, D and G-states. The extremely strong resonance due to the S-state, is due to the repulsive Coulomb interaction that introduces a barrier height and thus creates a pseudo-bound state.  
Considering each of the potential models in RK-5 algorithm for each of the $\ell$-channels, we have obtained corresponding best model parameters by minimizing the mean absolute percentage error (MAPE), given by
\begin{equation}
MAPE = \frac{1}{N}\sum_{i=1}^{N}\big|\frac{\delta_{i}^{expected}- \delta_{i}^{simulated}}{\delta_{i}^{expected}}\big|\times 100
\end{equation}
where $\delta_{i}^{expected}$ and $\delta_{i}^{simulated}$ are the expected and simulated scattering phase shifts respectively. This process of utilising all the available experimental SPS to determine the underlying interaction potential is akin to procedure of inverse scattering theory. Thus, each of the phenomenological models are in effect proposing different mathematical functions that guide in constructing the inverse potential for various $\ell$-channels \cite{37}. 

\begin{figure}
    \centering
    \includegraphics[scale=0.40]{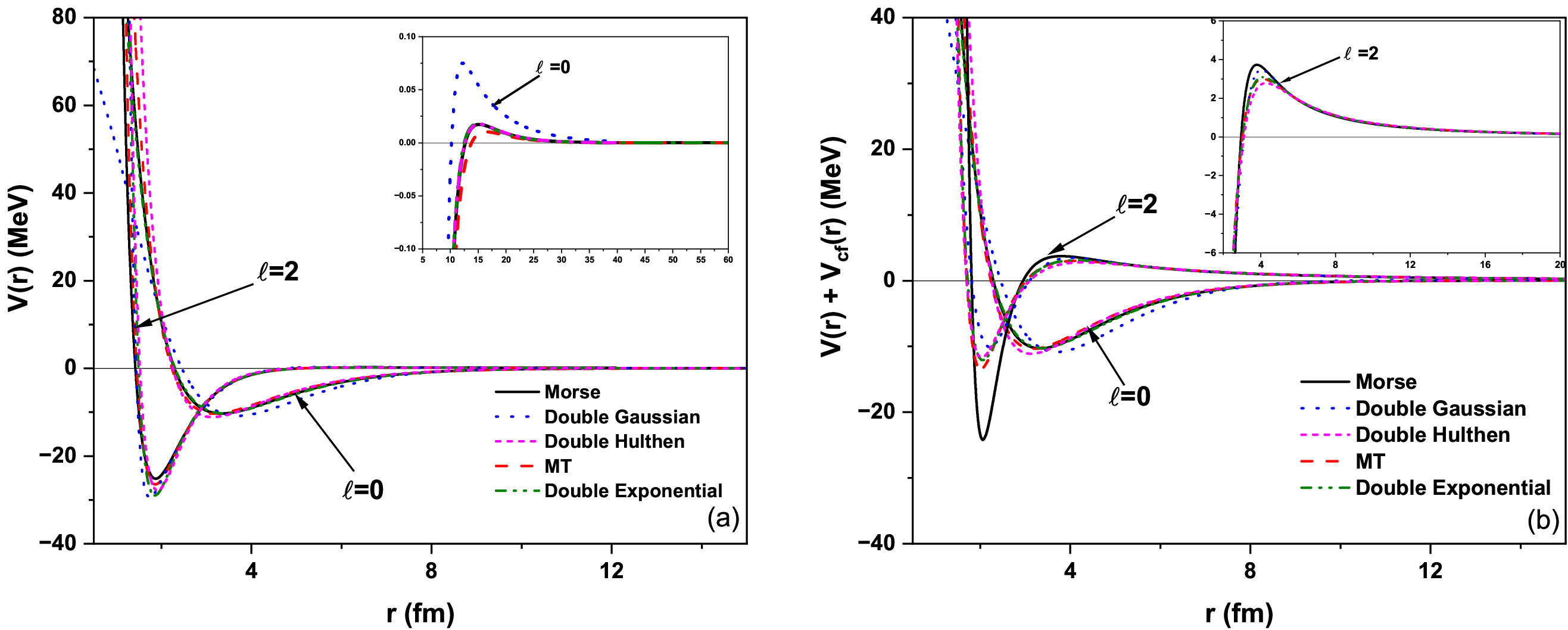}
    \caption{Interaction Potential without and with centrifugal term  $\ell$ = 0 and 2.}
    \label{fig1}
\end{figure}

Initially, we have treated the screening radius $a$ in atomic Hulthén potential as a free parameter and obtained the optimised parameters by integrating the phase equation to large distance of about 40 fm. This is not the case in ref \cite{10}, where in the erf() function was cutoff at about 6 fm to obtain the optimised parameters. The optimised parameters for $\ell=0, 2$ along with respective MAPE values are shown in Table 1. We did not show the optimised parameters for $\ell=4$, as the number of experimental data points available for this channel are only 4, whereas the number of parameters in case of double Hulthén, double Gaussian and double exponential is 5. This implies that the number of equations to be solved is less than the number of unknowns and the system is under-determined.
\begin{table}[h]
\caption{Model parameters of different mathematical functions for $\ell$ = 0 , 2 and 4 with screening radius 'a' as free parameter.}
\begin{tabular}{@{}|ccccc|@{}}
\hline
Mathematical Function   & $\ell$ & Optimized                 & Screening  & MAPE  \\
Model Parameters  &  &  Parameters                 & radius (a) &  \\
\hline
Morse                   & 0 & (10.90, 3.31, 1.52)          & 4.77               & 1.5 \\
($D_0$, $r_m$, $a_m$)              & 2 & (40.26, 2.02, 0.41)          & 3.31               & 2.0 \\ \hline
                        &   &                                  &                      &        \\
Double Gaussian         & 0 & (28.81, 97.46, 0.23, 0.51)  & 7.01               & 0.9  \\
($V_a$, $V_r$, $\mu_a$, $\mu_r$)            & 2 & (193.58, 499.6, 0.58, 0.86) & 3.55               & 2.4 \\  \hline
                        &   &                                  &                      &        \\                        

Double Hulthén          & 0 & (58.48, 44.35, 0.99, 0.36)    & 4.82               & 1.7   \\
($S_{\ell1}$, $S_{\ell2}$, $\beta$, $\alpha$)   & 2 & (1623.35, 1494.35, 3.74, 2.07) & 4.81             & 3.7 \\  \hline
                        &   &                                  &                      &        \\
MT                      & 0 & (1335.69, 443.49, 0.50)      & 4.54                & 1.7 \\
($V_R$, $V_A$, $\mu$)               & 2 & (28771.21, 264.31, 1.54)    & 4.39               & 3.3 \\  \hline
                        &   &                                  &                      &        \\
Double Exponential      & 0 & (78.65, 423.76, 1.22, 0.68)  & 4.69               & 1.4  \\
($A$, $B$, $\alpha_1$, $\alpha_2$)`     & 2 & (1994.15, 375.95, 3.07, 1.99)  & 4.13               & 3.1 \\ \hline
\end{tabular}
\label{tab1}
\end{table}
From Table 1, it is evident that the screening radius for  $\ell$ = 0 is greater than that for $\ell$ = 2. Therefore, we conclude that as the angular momentum ($\ell$) increases, the value of 'a' should decrease. 
One can observe that MAPE convergences to values between 1 to 2 $\%$ for $\ell=0$ and to between 2 to 4 $\%$ for $\ell=2$. The potential plots for these two channels without and with centrifugal term added are shown in Fig. \ref{fig1} (a) and (b) respectively. The inset of Fig \ref{fig1}(a) shows the centrifugal barrier height of $\ell=0$ for various model potentials and it is seen that all of them are not high enough to make the S-state to be a pseudo-bound state. In Fig. \ref{fig1}(b), the inset shows the centrifugal barrier heights of $\ell=2$, all of which are varying from 2.5 to 3.5 MeV. Even though the barrier heights are close to 3 MeV as one would expect from the observed resonance of $\ell=2$, the depths of the potential after adding the centrifugal term are not shallower than that of $\ell=0$.  All these observations made us realise that the optimised potentials are not physically realistic interactions. Hence, we have reoptimised the parameters to ensure that the following conditions are met:
\begin{enumerate}
\item The height of the Coulomb barrier for the S state is equal to or near the pseudobound state.
\item When the centrifugal term is added, the potential depth for the D state is lower than that for the S state.
\item The heights of the Coulomb barrier for the D and G states are near to their observed resonance energies respectively.
\end{enumerate}  

%From the inset of fig \ref{fig1}(a) , we observed that with these optimized parameters, the pseudo-bound state does not occur for $\ell$ = 0. Among the phenomenological models investigated, the Double Gaussian  yielded the highest value for the Coulomb barrier height, reaching around 0.07 MeV. Interestingly, the other phenomenological models exhibited barrier heights that were lower than that of the Double Gaussian function. From fig \ref{fig1}(b) , it is evident that when we incorporate the centrifugal term into the obtained interaction potentials, the depth for $
%\ell$ = 2 does not decrease compared to that for $\ell$ = 0. Hence, we can conclude that the optimized model parameters, fail to accurately explain the physical implications of $\alpha - \alpha$ scattering.
In the second iteration, we have chosen various values of the screening radius 'a' and examined its impact in elucidating $\alpha- \alpha$ scattering. To achieve this, we started increasing $a$ value and observed that the Coulomb barrier height kept increasing and so also the MAPE values. The obtained potential depth, barrier height and corresponding MAPE values have been compiled for values of $a$ from 10 to 25 fm in steps of 5 fm for $\ell=0$ S-state in table \ref{tab2}. 
Overall the trend is that while the depth of the potential decreases with increasing $a$ except for double gaussian, the barrier height increases close to expected 0.1 MeV. Similarly for D-state, $a$ values were increased in steps of 1 fm from 4 fm onwards. It was found that up to 6 fm, the potential depth remained higher than that of S-state and only at 8 to 9 fm, the depths became shallower except for the double exponential function. The barrier height keeps decreasing with increasing $a$ and MAPE steadily increases as well. Finally, now that screening parameter $a$ is being fixed for a particular optimisation run, we could obtain the parameters for $\ell=4$ as well. It was observed that for higher $\ell$ values the screening parameter reduces. Hence, the values of $a$ were started at even smaller value and more fine tuned by varying only in steps of 0.5 fm now, from 3 to 4.5 fm. The barrier height keeps decreasing with increasing screening radius as in case of D-state. On the other hand, MAPE values tend to decrease for MT and double Hulthén, increase in case of double gaussian and reach a minima in cases of Morse and double exponential for some in between $a$ values. While double exponential gives best MAPE of 0.1 at a = 4fm, Morse has best value of 0.5 at 3.5 fm. One can choose different sets of $a$ values for each of the 5 model potentials as given in Table \ref{tab3} and compare the respective interaction potentials for S, D and G states. 

\begin{table}[h]
\caption{Potential depth, Barrier height and Mape at different screening radius ('a') for $\ell$ = 0, 2 and 4.}
\scalebox{0.8}{
\setlength{\tabcolsep}{8pt} % Default value: 6pt
\renewcommand{\arraystretch}{1.2}
\begin{tabular}{@{}|l|l|l|l|l|@{}}
\hline
\multicolumn{5}{l}{~~~~~~~~~~~~~~~~~~~~~~~~~~~~~~~~~~~~~~~~~~~~~~~~~~~~~~~~~~~~~~~~~~~~~~~~~~~~$\ell=0$}                                                                                                            \\ \hline
MF/a               & 10fm                      & 15fm                      & 20fm                       & 25fm                      \\ \hline
Morse              & {[}-10.85, 0.07, 1.6{]} & {[}-10.43, 0.09, 1.6{]} & {[}-10.37, 0.11, 1.6{]}  & {[}-10.28, 0.11, 1.6{]} \\
Double Gaussian    & {[}-11.32, 0.11, 0.8{]} & {[}-11.84, 0.14, 0.9{]} & {[}-12.07, 0.16, 0.9{]}  & {[}-12.17, 0.17, 0.9{]} \\
Double Hulthén     & {[}-11.23, 0.06, 1.8{]} & {[}-11.33, 0.09,
1.8{]} & {[}-11.52, 0.11, 2.1{]}  & {[}-11.50, 0.12, 2.1{]} \\
MT                 & {[}-10.95, 0.06, 1.7{]} & {[}-10.23, 0.08, 1.7{]} & {[}-10.09, 0.09, 1.7{]}  & {[}-9.95, 0.10, 1.6{]}  \\
Double Exponential & {[}-10.73, 0.07, 1.4{]} & {[}-10.46, 0.09, 1.5{]} & {[}-10.40, 0.11, 1.5{]}  & {[}-10.51, 0.13, 1.5{]} \\\hline
\multicolumn{5}{l}{~~~~~~~~~~~~~~~~~~~~~~~~~~~~~~~~~~~~~~~~~~~~~~~~~~~~~~~~~~~~~~~~~~~~~~~~~~~~$\ell=2$} \\ \hline
MF/a               & 6fm                       & 7fm                       & 8fm                        & 9fm                       \\ \hline
Morse              & {[}-10.82, 2.81, 3.8{]} & {[}-9.56, 2.69, 4.3{]}  & {[}-8.82, 2.63, 4.5{]}    & {[}-8.51, 2.56, 4.3{]}  \\
Double Gaussian    & {[}-12.98, 2.96, 4.2{]} & {[}-11.28, 2.87, 4.6{]} & {[}-10.23, 2.82, 4.9{]}  &  {[}-10.08, 2.78, 5.2{]}                         \\
Double Hulthén     & {[}-11.77, 2.74, 3.8{]} & {[}-10.75, 2.63, 4.3{]} & {[}-10.04, 2.55, 4.6{]}  & {[}-9.59, 2.49, 4.8{]}  \\
MT                 & {[}-11.85, 2.85, 3.8{]}  & {[}-10.56, 2.69, 4.2{]} & {[}-9.80, 2.61, 4.5{]}   & {[}-9.34, 2.55, 4.7{]}  \\
Double Exponential & {[}-14.29, 2.81, 3.7{]} & {[}-13.53, 2.68, 4.2{]} & {[}-11.77, 2.60 ,4.6{]} & {[}-11.57, 2.55, 4.7{]} \\
 \hline
\multicolumn{5}{l}{~~~~~~~~~~~~~~~~~~~~~~~~~~~~~~~~~~~~~~~~~~~~~~~~~~~~~~~~~~~~~~~~~~~~~~~~~~~~$\ell=4$}                                                                                                            \\ \hline
MF/a               & 3fm                       & 3.5fm                     & 4fm                        & 5fm                       \\ \hline
Morse              & {[}-0.02, 10.98, 1.3{]} & {[}0.06, 10.80, 0.5{]}  & {[}0.06, 10.67, 0.7{]}   & {[}0.06, 10.44, 1.2{]}  \\
Double Gaussian    & {[}0.13, 9.79, 2.7{]}   & {[}0.13, 9.61, 3.3{]}   & {[}0.13, 9.53, 3.8{]}    & {[}0.13, 9.32, 4.6{]}   \\
Double Hulthén     & {[}-2.67, 11.56, 3.7{]} & {[}-1.83, 11.37, 2.9{]} & {[}-1.11, 11.22, 2.3{]}  & {[}0.13, 10.95, 1.2{]}  \\
MT                 & {[}-3.13, 11.59, 3.4{]} & {[}-2.312, 11.40, 2.6{]} & {[}-1.61, 11.24, 2.0{]}  & {[}-0.27, 10.95, 1.1{]} \\
Double Exponential & {[}0.03, 10.74, 0.7{]}  & {[}0.03, 10.67, 0.4{]}  & {[}0.03, 10.71, 0.1{]}   & {[}0.03, 10.56, 1.1{]}  \\
 \hline
\end{tabular}
}
\label{tab2}
\end{table}
\begin{table}[h]
\caption{Optimised Model parameters of different mathematical functions for $\ell$ = 0, 2 and 4. The screening radius 'a' is shown in bold.}
\scalebox{0.65}{
\setlength{\tabcolsep}{7pt} % Default value: 6pt
\renewcommand{\arraystretch}{1.0}
\begin{tabular}{@{}|ccccc|@{}}
\hline
Mathematical   Function & Model Parameters            & $\ell$ =0                                & $\ell$=2                                 & $\ell$=4                              \\ \hline
Morse                   & ($D_0$, $r_m$, $a_m$, \textbf{a})               & (11.18, 3.42, 1,63, \textbf{15.0})          & (27.06, 1.89, 0.63, \textbf{7.0})            & (241.66, 0.37, 0.74, \textbf{3.5})        \\
                        & MAPE                        & 1.6                                & 4.3                                 & 0.5                              \\ \hline
                        &                             &                                    &                                     &                                  \\
Double Gaussian         &($V_a$, $V_r$, $\mu_a$, $\mu_r$, \textbf{a})             & (42.79, 98.36, 0.24, 0.44,\textbf{ 20.0})    & (68.93 , 36102.08, 0.48, 1.78, \textbf{8.0}) & (128.68, 1.24, 0.49, 4.46 , \textbf{0.5}) \\
                        & MAPE                        & 0.9                                & 4.6                                 & 0.5                              \\ \hline
                        &                             &                                    &                                     &                                  \\
Double Hulthén          & ($S_{\ell1}$, $S_{\ell2}$, $\beta$, $\alpha$, \textbf{a}) & (48.54, 35.67, 1.49, 0.90, \textbf{15.0})   & (1065.54, 963.88, 2.10, 0.54, \textbf{7.0})  & (48.17, 1.33, 5.36, 4.14, \textbf{5.0})   \\
                        & MAPE                        & 1.9                                & 4.3                                 & 1.2                              \\ \hline
                        &                             &                                    &                                     &                                  \\
MT                      & ($V_R$, $V_A$, $\mu$,\textbf{ a})               & (1080.63, 408.32, 0.43, \textbf{20.0})      & (8284,89, 1330.72, 1.27, \textbf{8.0})       & (958.32, 857.18, 1.01, \textbf{5.0})      \\
                        & MAPE                        & 1.7                                & 4.5                                 & 1.1                              \\ \hline
                        &                             &                                    &                                     &                                  \\
Double Exponential      & ($A$, $B$, $\alpha_1$, $\alpha_2$, \textbf{a})      & (117.14 , 80.59, 0.89, 0.73, \textbf{25.0}) & (8597.48, 52.55, 4.86, 1.39, \textbf{9.0})   & (55.51, 68.62, 3.08, 1.32, \textbf{4.0})  \\
                        & MAPE                        & 1.5                                & 4.7                                 & 0.1                              \\ \hline
\end{tabular}
}
\label{tab3}
\end{table}

\begin{figure}[h]
    \centering
    \includegraphics[scale=0.40]{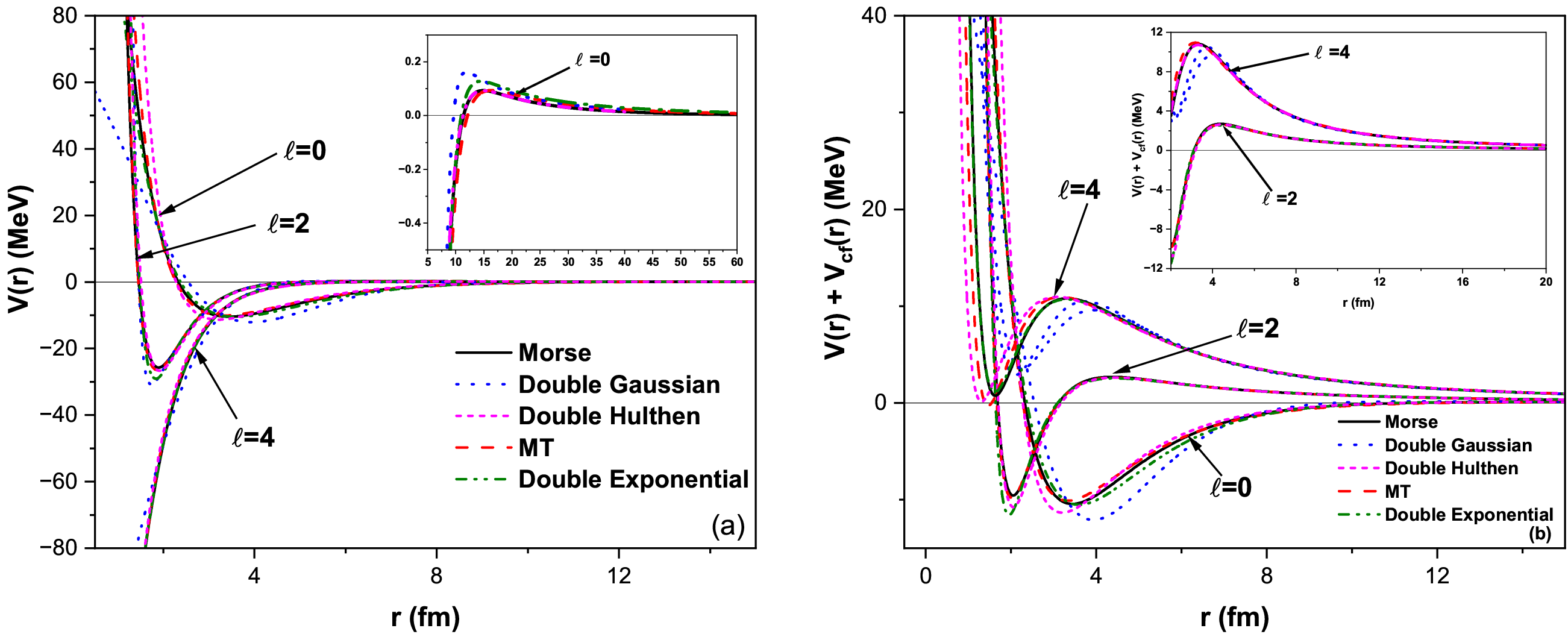}
    \caption{Interaction Potentials without and with centrifugal potential  $\ell$ = 0 , 2 and 4.}
    \label{fig2}
\end{figure}

The interaction potentials, with and without the centrifugal term, is depicted in fig. \ref{fig2}. From the inset of fig. \ref{fig2}(a), it is evident that pseudo-bound states are obtained for all phenomenological models. Additionally, fig. \ref{fig2}(b) reveals that the inclusion of the centrifugal term causes the depth of the potential to be lower for $\ell$ = 2 and 4 compared to $\ell$ = 0, for all models except the Double Exponential model. Based on all these comparative observations, one can easily that the inverse potentials obtained using any of the chosen mathematical models are exactly same, with negligible variations, as they all converge to give mean absolute percentage errors to within about $1\%$. So, even though many different mathematical functions have been proposed over the years, they all guide the process of constructing inverse potentials in exactly same manner. 

\begin{figure}[h]
    \centering
    \includegraphics[scale=0.5]{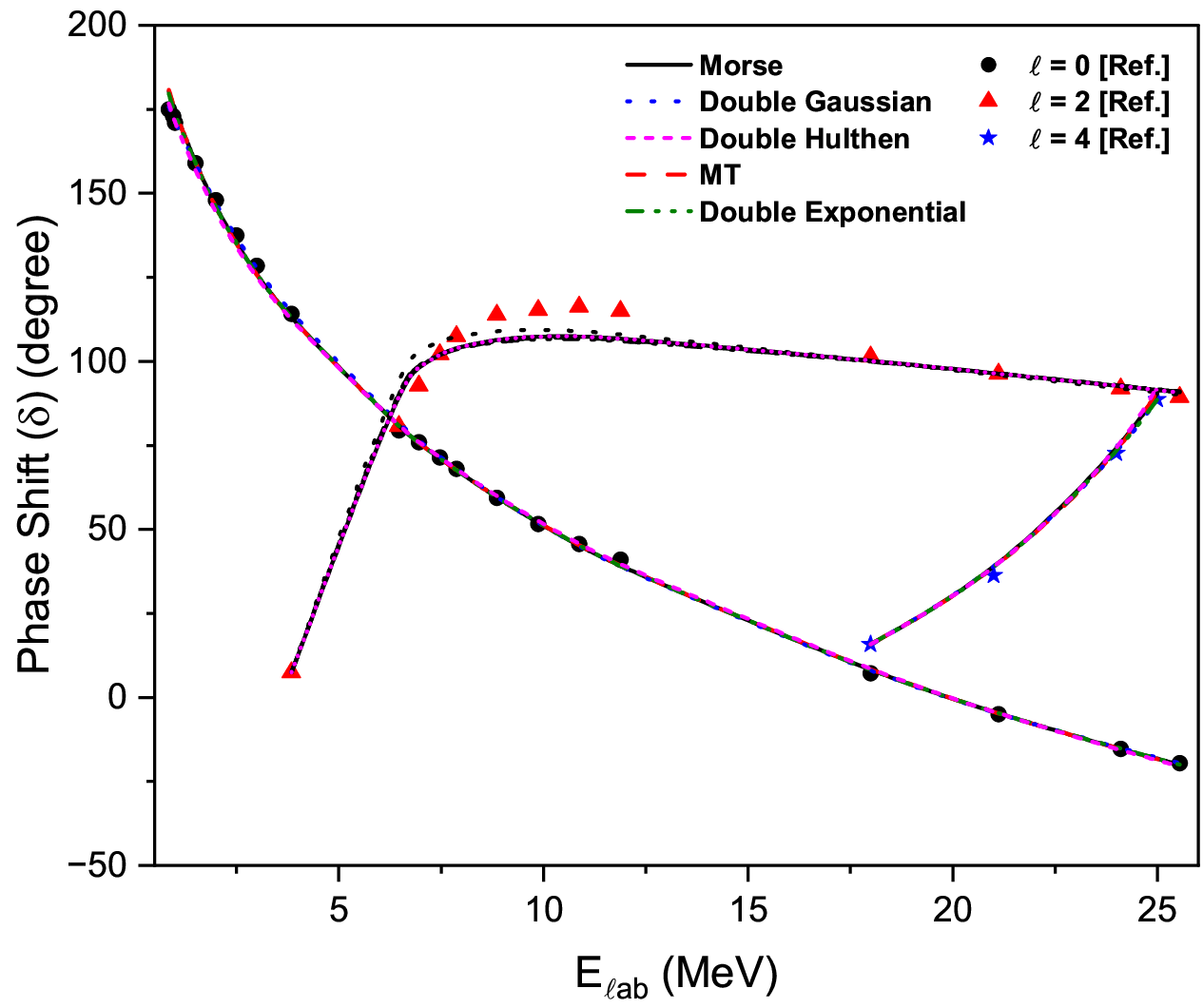}
    \caption{Obtained Scattering phase shifts for  $\ell$ = 0 , 2 and 4 along with expected phase shifts given in Ref. \cite{10}}
    \label{fig3}
\end{figure}

One might think that this might be due to global optimisation algorithm which seems to always converge to similar shape for the inverse potentials.
So, to test this, we have considered only as many experimental data points as per the number of model parameters, so that the equations are neither under-determined nor over-determined. That is, we have obtained interaction potentials, for three parameter Morse and MT potentials, by considering the following energies for 
each of the partial waves during optimisation: 
\begin{enumerate}
\item $\ell = 0$: E=[0.85 MeV, 9.88 MeV, 25.55 MeV]
\item $\ell = 2$: E=[3.84 MeV, 7.47 MeV, 25.55 MeV]
\item $\ell = 4$: E=[18 MeV, 21.13 MeV, 25.55 MeV]
\end{enumerate}
Similarly, for mathematical functions with four parameters, such as double Gaussian, double Hulthén, and double exponential potentials, we have extended our analysis by adding one additional energy point for each $\ell$ value. They are $2.5 MeV, 18 MeV, 24.11 MeV$ for $\ell = 0$, $2$, and $4$ respectively. The obtained parameters for S, D and G-states considering each of the model potentials are compiled in Table \ref{tab4}. It is evident that the results are comparable to those obtained from the global optimization algorithm. Even the mean absolute percentage errors obtained are only slightly higher than those obtained using GOA.
\begin{table}[h]
\caption{Optimized Model Parameters of Interaction Potential for $\ell$ = $0$, $2$, and $4$ by taking number of data points equal to number of model parameters.}
\scalebox{0.65}{
\setlength{\tabcolsep}{7pt} % Default value: 6pt
\renewcommand{\arraystretch}{1.0}
\begin{tabular}{@{}|ccccc|@{}}
\hline
Mathematical   Function & Model Parameters            & $\ell$ =0                                & $\ell$=2                                 & $\ell$=4                              \\ \hline
Morse                   & ($D_0$, $r_m$, $a_m$, \textbf{a})             & (11.46, 3.35, 1.58, 15.0)          & (27.04, 1.92, 0.62, 7.0)           & (214.16, 0.47, 0.73, 3.5)        \\
                        & MAPE                      & 2.3                                & 4.5                                & 1.0                                \\ \hline
                        &                           &                                    &                                    &                                  \\
Double Gaussian         &($V_a$, $V_r$, $\mu_a$, $\mu_r$, \textbf{a})                 & (50.77, 100.42, 0.24, 0.41, 20.0 ) & (53.97, 15558.85, 0.45, 1.80, 8.0) & (128.68, 1.24, 0.49, 4.46 , 0.5) \\
                        & MAPE                      & 1.4                                & 5.2                                & 0.5                              \\ \hline
                        &                           &                                    &                                    &                                  \\
Double Hulthén          & ($S_{\ell1}$, $S_{\ell2}$, $\beta$, $\alpha$, \textbf{a}) & (49.74, 36.61, 3.42, 2.82, 15.0)   & (824.72, 737.18, 1.56, 0.12, 7.0)  & (48.17, 1.33, 5.36, 4.14, 5.0)   \\
                        & MAPE                      & 3.2                                & 5.4                                & 1.2                              \\  \hline
                        &                           &                                    &                                    &                                  \\
MT                      & ($V_R$, $V_A$, $\mu$,\textbf{ a})                      & (1189.46, 432.15, 0.46, 20.0)      & (10737.19, 1527.78, 1.31, 8.0)     & (952.66, 851.72, 1.01, 5.0)      \\
                        & MAPE                      & 2.7                                & 4.7                                & 1.3                              \\ \hline
                        &                           &                                    &                                    &                                  \\
Double Exponential      & ($A$, $B$, $\alpha_1$, $\alpha_2$, \textbf{a})        & (73.73, 14.53, 1.31, 0.59, 25.0)   & (6381.90, 62.53, 4.50, 1.44, 9.0)  & (55.51, 68.62, 3.08, 1.32, 4.0)  \\
                        & MAPE                      & 2.5                                & 4.9                                & 0.1                             \\ \hline
\end{tabular}
}
\label{tab4}
\end{table}
The obtained SPS for S, D, and G states are shown in Figure \ref{fig3}. The obtained scattering phase shifts follow the same trend as the expected ones \cite{10} for $\ell$ = 0 and 4. However, there are slight discrepancies from a lab energy of 7.88 MeV to 11.88 MeV for $\ell$ = 2. Therefore, we can conclude that the atomic Hulthén as a screened Coulomb potential works well for the S and G states, but it is not as effective in capturing the peak that appears in the SPS of the D state, when we use phase function method to calculate SPS. One can also conclude that all the mathematical functions are more or less equally effective in guiding the construction of inverse potentials for all the $\ell$-channels.
\section{Conclusions}
The inverse potentials for alpha-alpha scattering have been constructed by considering various successful models proposed for nuclear interactions, such as Morse, double Gaussian, double Hulthén Malfliet-Tjon and Double exponential functions with atomic Hulthén as ansatz for screened Coulomb interaction. The model parameters have been optimised using a global optimisation algorithm \cite{38} which minimises mean absolute percentage error between the obtained scattering phase shifts from phase function method and the experimental data. On comparison of the resultant inverse potentials, one can conclude that all the mathematical models agree with each other to within small variations with almost similar mean absolute percentage errors. Since, inverse potential approach utilises all available experimental data, they provide a globally optimal solution which might become data dependent. Hence, we have also performed optimisation by considering only as many experimental data points as the number of model parameters. This procedure also lead to similar interaction potentials to those obtained using global optimisation. So, it is reasonable to conclude that all mathematical functions considered only serve to guide the process of obtaining the interaction potential and are not unique. This is going to be true for any potential, as long as, it has the basic required features of any two body interaction, which are repulsion at short distances, attractive nature for intermediate distances and exponentially falling of tail for large distances.
\section*{Acknowledgments}
A. Awasthi acknowledges financial support provided by Department of Science and Technology
(DST), Government of India vide Grant No. DST/INSPIRE Fellowship/2020/IF200538.   \\
 
\textbf{Author Declaration} The authors declare that they have no conflict of interest.

\end{document}